\documentclass{article}

\usepackage{doc}

\usepackage{amsthm}
\usepackage{hyperref}

\theoremstyle{definition}

\begin{document}

\author{Marijn J.H. Heule}

\title{The DRAT format and DRAT-trim checker}

\maketitle

The proof checker DRAT-trim~\cite{Wetzler14} can be used to check whether a
propositional formula in the DIMACS format is unsatisfiable. Given
a propositional formula and a clausal proof, DRAT-trim validates
that the proof is a certificate of unsatisfiability of the formula.
Clausal proofs should be in the DRAT format which is used to validate
the results of the SAT competitions. The format is defined below.

The DRAT format and the corresponding checker DRAT-trim are designed
based on the following goals: 1) It should be easy to emit clausal
proofs to ensure that many SAT solvers will support it; 2) proofs should
be compact to reduce the overhead of writing the proof and to make sure
that the proof can be stored; 3) proof validation should be efficient;
and 4) all techniques used in state-of-the-art SAT solvers should be
expressible in the format. Regarding the last point, several techniques
used in state-of-the-art SAT solvers cannot be expressed using resolution.
The DRAT format is therefore a generalization of Extended Resolution.

In short, each step in a clausal proof is either the addition or the
deletion of a clause. Each clause addition step should preserve
satisfiability, which should be computable in polynomial time. The
polynomial time checking procedure is described in detail below.
Clause deletion steps are not checked, because they trivially
preserve satisfiability. The main reason to include clause deletion
steps in proofs is to reduce the computational costs to validate proofs.

\section*{Syntax of Input Files}

The DRAT-trim checker requires two input files: a CNF formula in the
DIMACS format and a clausal proof in the DRAT format. Both formats
are defined below.

The syntax of a CNF formula in DIMACS format is:

\begin{verbatim}
  <formula>   = { <comment> } "p cnf " <var-max> " " <num-cls> "\n" <cnf>
  <cnf>       = { <comment> | <clause> }
  <comment>   = "c " <anything> "\n"
  <clause>    = { <blank> }{ <lit> } "0"
  <lit>       = <pos> <blank> | <neg> <blank>
  <pos>       = "1" | "2" | ... | <max-idx>
  <neg>       = "-" <pos>
  <blank>     = " " | "\n" | "\t"
\end{verbatim}

\noindent
where {\tt |} means choice, {\tt \{ $\dots$ \}} is equivalent to the Kleene star
operation (that is a finite number of repetitions including 0) and
{\tt <max-idx>} is $2^{31} - 1$.

In the first line of a DIMCAS formula, {\tt <var-max>} should be at least
as high as the largest {\tt <pos>} used in the formula, while {\tt <num-cls>}
should equal the number of clauses in the formula. In case there
exists a literal {\tt <pos>} or {\tt "-" <pos>}, such that <pos> is larger than
{\tt <var-max>}, then the formula is invalid. The formula is also invalid
if the number of clauses is not equal to {\tt <num-cls>}. 

The DIMCAS format is the default format used in SAT solvers and
has been used as the input format by all SAT Competitions.

The syntax of a clausal proof in DRAT format is as follows:

\begin{verbatim}
  <proof>     = { <comment> | <clause> | "d" <blank> <clause> }
  <comment>   = "c " <anything> "\n"
  <clause>    = { <blank> }{ <lit> } "0"
  <lit>       = <pos> <blank> | <neg> <blank>
  <pos>       = "1" | "2" | .... | <max-idx>
  <neg>       = "-" <pos>
  <blank>     = " " | "\n" | "\t"
\end{verbatim}

Notice that the syntax of the DIMACS and DRAT formats is equivalent
for {\tt <clause>}, {\tt <lit>}, {\tt <pos>}, and {\tt <neg>}.

However, DIMACS and DRAT files differ in their interpretation.
Formulas in the DIMACS files are multi-sets of clauses. Hence,
swapping clauses in a DIMACS file preserves the interpretation.
On the other hand, DRAT files are a sequence of clause addition
and deletion steps. As a consequence, the order of the clauses
influences the proof.

A clausal proof is valid with respect to a given formula, if
1) all clause addition steps are valid (details are below); and
2) if the proof contains the empty clause ({\tt "0"}).

The empty clause typically is the last clause of the proof file,
because all lines after the empty clause are redundant and ignored.

\section*{Clause Syntax Restrictions}

There are two restrictions regarding clauses in both the DIMACS
and the DRAT formats. The first restriction is that no clause can
be a tautology, i.e., contains a complementary set of literals.
Hence if a clause contains {\tt <lit>} it cannot contain {\tt "-" <lit>} as well.
The second restriction is that clauses cannot contain duplicate
literals. On the other hand, formulas are allowed to have
duplicate clauses. Hence formulas are multi-sets of clauses.

\section*{Clause Addition Preconditions}

A clause with only one literal is called a unit clause. Checking
whether a clause is redundant with respect to a CNF formula is
computed via Unit Clause Propagation (UCP). UCP works as follows:
For each unit clause $(l)$ all literal occurrences of $\bar l$ are removed
from the formula. Notice that this can result in new unit clauses.
UCP terminate when either no literals can be removed or when it
results in a conflict, i.e., all literals in a clause have been
removed.

Let $C$ be a clause. $\overline{C}$ denotes the negation of a clause, which
is a conjunction of all negated literals in $C$. A clause C has the
redundancy property Asymmetric Tautology (AT) with respect to a CNF
formula $F$ iff UCP on $F \land (\overline C)$ results in a conflict. The core
redundancy property used in the DRAT format is Resolution Asymmetric
Tautology (RAT). A clause $C$ has the RAT property with respect to a
CNF formula $F$ if there exists a literal $l \in C$ such that for all
clauses $D$ in $F$ with $\lnot l \in D$, the clause $C \lor (D \setminus \{\lnot l\})$
has the property AT with respect to $F$. Notice that RAT property is a
generalization of the AT property.

The DRAT proof checking works as follows. Let $F$ be the input formula
and $P$ be the clausal proof. At each step $i$, the formula is modified.
The initial state is: $F_{0} = F$. At step $i > 0$, the $i^{th}$ line of $P$ is
read. If the line has the prefix {\tt d}, then the clause C described on
that line is removed: $F_{i} = F_{i-1} \setminus \{C\}$. Otherwise, if there is
no prefix, then C must have the RAT property with respect to formula
$F_{i-1}$. This must be validated. Recall that the RAT property requires
a pivot literal $l$. In the DRAT formula it is expected that the first
literal in $C$ is the pivot. If the RAT property can be validated, then
the clause is added to the formula: $F_{i} = F_{i-1} \land C$. If the
validation fails, then the proof is invalid.

The empty clause at the end of the proof should have the AT property
as it does not have a first literal.

\section*{Clause Deletion Details}

The only expectation of a clause deletion step is that the to-be-deleted
clause is present in the current formula. Clause deletion steps are ignored
if the clause is not present in the current formula. DRAT-trim will print
a warning to notify the user about such unexpected steps.

A clause deletion step removes a single clause (if the clause is present).
Since formulas are multi-sets of clauses, duplicate clauses may exist.
In order to remove all $k$ copies of a clause, $k$ clause deletion steps are
required in a clausal proof. If case a duplicate clause is deleted, then
DRAT-trim removes an arbitrary copy.

DRAT-trim ignores deletion steps of unit clauses by design. This decision
facilities a more efficient implementation and reduces the complexity of the
checker. The main reason to add deletion information to a clausal proof is
to reduce the computation costs to validate that proof. However, deletion
of unit clauses has the opposite effect. Notice that ignoring deletion steps
of unit clauses can turn a valid DRAT proof into an invalid one --- and the
other way around. DRAT-trim therefore prints a warning statement to inform
the user about such a modification.

\section*{Example}

Consider the following CNF formula in DIMACS format (with spacing to
increase readability):
\begin{verbatim}
   p cnf 4 8
    1  2 -3 0
   -1 -2  3 0
    2  3 -4 0
   -2 -3  4 0
   -1 -3 -4 0
    1  3  4 0
   -1  2  4 0
    1 -2 -4 0
\end{verbatim}

\noindent
A valid DRAT proof for the above formula is:

\begin{verbatim}
         -1 0
  d -1 2  4 0
          2 0
            0
\end{verbatim}

The first step is to validate that the first clause in the proof {\tt -1 0}
has the RAT property with respect to $F_{0}$ (the input formula). This RAT
check can be partitioned into three AT checks: there are three clauses
containing the literal {\tt 1}: {\tt1 2 -3 0}, {\tt 1 3 4 0} and {\tt 1 -2 -4 0}.
Following the description above, the RAT checks requires checking the AT
check for the clauses {\tt -1 2 -3 0}, {\tt -1 3 4 0} and {\tt -1 -2 -4 0}. For the
first AT check, UCP on the formula $F_{0} \land (1) \land (\lnot 2) \land (3)$ should result
the empty clause. First, clause {\tt -1 -3 -4 0} is reduced to the unit $(\lnot 4)$.
Now the clause {\tt -1 2 4 0} is reduced to the empty clause.

The RAT check of {\tt -1 0} succeeds. Now {\tt-1 0} will be added to the formula:
$F_{1} = F_{0} \land (\lnot 1)$.

The second step is the removal of the clause {\tt d -1 2 4 0}. Deletion steps
are not checked. This results in $F_{2} = F_{1} \setminus (\lnot 1 \lor 2 \lor 4)$.

\section*{Remarks on Efficient Proof Validation}

In order to efficiently validate a DRAT proof, several optimizations to the
checking algorithm are required. Examples of such optimizations are backward
checking and core first unit propagation. Details about these techniques are
described in~\cite{Heule:2013:trim}.

\section{Binary DRAT Format}

This section describes the "binary format" for input proof files.

\subsection*{Mapping DIMACS Literals to Unsigned Integers}

The first step of the binary encoding is mapping literals in the DIMACS format
to unsigned integers. The following mapping function is used:

\[
map(l) := (l > 0)~?~2 \cdot l : -2 \cdot l + 1. 
\]

The mapping for some DIMACS literals are shown
below.

\begin{verbatim}
     DIMACS literals   unsigned integers

                 -63     127
                 129     258
               -8191   16383
               -8193   16387
\end{verbatim}

\subsection*{Variable-Byte Encoding of Unsigned Integers}

Assume that {\tt w0}, $\dots$, {\tt wi} are 7-bit words, {\tt w1} to {\tt wi} all non zero and the
unsigned number {\tt x} can be represented as

\begin{verbatim}
    x = w0 + 2^7*w1 + 2^14*w2 + 2^(7*i)*wi
\end{verbatim}

The variable-byte encoding of DRAT (also used in AIGER) is the sequence of $i$
bytes {\tt b0}, ..., {\tt bi}:

\begin{verbatim}
    1w0, 1w1, 1w2, ..., 0wi
\end{verbatim}

The MSB of a byte in this sequence signals whether this byte is the last byte in
the sequence, or whether there are still more bytes to follow. Here are some examples:

\begin{verbatim}
    unsigned integer   byte sequence of encoding (in hexadecimal)

                   x   b0 b1 b2 b3 b4

                   0   00
                   1   01
    2^7-1    =   127   7f
    2^7      =   128   80 01
    2^8  + 2 =   258   82 02
    2^14 - 1 = 16383   ff 7f
    2^14 + 3 = 16387   83 80 01
    2^28 - 1           ff ff ff 7f
    2^28 + 7           87 80 80 80 01
\end{verbatim}

\subsection*{Bringing it together}

In the binary DRAT format, each clause consists of at least two bytes. The first
byte expresses whether the lemma is added (character {\tt a} or {\tt 61} in hexadecimal) or
deleted (character {\tt d} or {\tt 64} in hexadecimal). The last byte of each lemma is the
zero byte (00 in hexadecimal). In between these two bytes, the literals of the
lemma are shown as unsigned integers in the variable-byte encoding.

In the example below, the plain DRAT format requires $26$ bytes (including the new
line characters and excluding the redundant spaces in the second lemma), while
the binary DRAT format of the same proof requires only $12$ bytes. Emitting proofs
in the binary format reduces the size on disk by approximately a factor of three
compared to the conventional (plain) DRAT format.

\begin{verbatim}
          plain DRAT      binary DRAT (in hexadecimal)

       d -63 -8193 0      64 7f 83 80 01 00 61 82 02 ff 7f 00
         129 -8191 0
\end{verbatim}

References:



\bibliographystyle{plain}
\bibliography{DRAT}

\end{document}